# Electrostatic actuation of silicon optomechanical resonators

Suresh Sridaran and Sunil A. Bhave
OxideMEMS Lab, Cornell University, Ithaca, NY, USA

**Optomechanical systems offer one of the most sensitive methods for detecting mechanical motion using shifts in the optical resonance frequency of the optomechanical resonator [1]. Presently, these systems are used for measuring mechanical thermal noise displacement [1,2,3,4] or mechanical motion actuated by optical forces [5,6,7,8,9]. Electrostatic capacitive actuation and detection have been shown previously for silicon micro electro mechanical resonators for application in filters and oscillators [10,11,12,13]. Here, we demonstrate monolithic integration of electrostatic capacitive actuation with optical sensing using silicon optomechanical disk resonators and waveguides. The electrically excited mechanical motion is observed as an optical intensity modulation when the input electrical signal is at a frequency of 235MHz corresponding to the radial vibrational mode of the silicon microdisk.**

Recent work on cavity optomechanics has shown resonators with mechanical displacement sensitivities close to the zero point motion of the mechanical modes [1]. The mechanical motion of these optomechanical resonators is encoded as phase shifts which can be extracted using homodyne detection [10] or by using the optical resonator to convert the phase changes into optical intensity modulation [6]. In addition to offering excellent displacement sensitivities, optical forces have been shown to exist in these optomechanical systems which can be used for a variety of applications including static motion of micro mechanical structures [8,9,10], setting up oscillations (heating) of the vibrational modes [5,6,7] and cooling of vibrational modes to achieve ground state[1,2,3,4]. The optomechanical systems can be either stand alone optomechanical resonators on chip which are interrogated by evanescent coupling from fiber taper that provide lower insertion loss [1,5,6,7,10] or on-chip systems which incorporate waveguides along with optomechanical resonators on the same chip [8,9].

Electrostatic capacitive actuation and detection is the main scheme of transduction used in Radio Frequency (RF) Micro Electro Mechanical Systems (MEMS) resonators [11,12,13,14]. The use of electrostatics is convenient as it enables direct integration with electronics used for processing RF signals. In electrostatic capacitive actuation, an electric potential is applied between a capacitor constituted by the electrodes surrounding the resonator and the resonator. This field across the capacitor leads to a force on the resonator. A combination of AC and DC potentials are used to obtain a force at a desired frequency. To sense mechanical displacement, the time varying capacitor between the resonator and sense electrodes leads to a current at the frequency of vibration.

In addition to the two schemes described above, in which the actuation and sensing are restricted to the respective domains, it is interesting to consider a crossover scheme where mechanical motion is excited electrically and sensed optically. An earlier work at using electrical actuation with optomechanical systems made use of a standalone optomechanical system with a gradient electrical force on a dielectric provided by an electrode off chip [15]. In this work, we present a method for driving an optomechanical resonator using the RF MEMS technique of electrostatic capacitive actuation and sensing of mechanical motion by using the optical intensity modulation at the output of an optomechanical resonator all integrated into a monolithic system fabricated on silicon on insulator (SOI) platform.

Figure 1 shows the schematic of the proposed electrically actuated optomechanical resonator consisting of two suspended microdisk resonator. The microdisk resonator on the left has electrodes around it to provide the electrostatic actuation force for driving mechanical motion. This serves as the driving disk. As the force is proportional to the square of potential across the actuation capacitor, the addition of DC bias results in a force at the RF frequency $\omega$ given by $F_{es}(\omega) = (dC/dg)V_{dc}V_{rf}$ where $(dC/dg)$ is the change in the actuation capacitance (C) for a small change in the electrode to resonator capacitor gap (g). This leads to a radial displacement at the mechanical resonant frequency corresponding to $\Delta r(\omega) = Q_{mech}F_{es}(\omega)/k_{eff}$, where $Q_{mech}$ is the mechanical quality factor and $k_{eff}$ is the effective spring constant of the radial vibrational mode. The disk on the right (colored gray) acts as the optomechanical resonator that senses mechanical motion as a phase shift which causes a variation in its optical resonance frequency. The sensitivity of the optomechanical resonator to radial displacement is obtained from the optical resonance condition $m\lambda_0 = 2\pi R n_{eff}$ where $m$ is an integer; $n_{eff}$ is the effective index obtained for the mode and $\lambda_0$ is free space wavelength. The displacement $\Delta r(\omega)$ gives a shift in optical resonance wavelength of $\Delta\lambda = \frac{\Delta r(\omega)}{R}\lambda_0$. For an optical resonance at 1550nm, a radial displacement of 0.5nm of a 10μm radius disk shifts the resonant wavelength by 77.5pm. This is observed as intensity modulation at frequency $\omega$ of a continuous wave laser input in a waveguide close to the resonator with wavelength corresponding to the half maximum point of optical resonance.

The drive and sense resonators are mechanically coupled with a coupling beam to transfer the vibrations. This prevents the presence of actuation electrodes from degrading the optical quality factor of the optomechanical resonator. This geometry of using mechanically coupled disks is similar to that used in electrostatic capacitive actuation to reduce signal feed through [16]. The coupling beam splits the radial breathing mode of the disk into two modes, one where the motion in the two disks is in phase and one where the motion is out of phase. The combination of the two mechanical modes and the coupling beam result in a filter response to the electrical input with the bandwidth being tunable by the characteristics of the coupling beam.

Silicon-on-Insulator (SOI) has been chosen as the system for implementation of the modulator. This is due to the ability to obtain high $Q_{opt}$ optical resonators [17] and high $Q_{mech}$ mechanical resonators [18] with electrical actuation on SOI. A scanning electron microscope image of a device fabricated using the steps described in the methods section is shown in figure 2. The radii of the disks are 10μm which gives a mechanical resonance frequency of 242MHz. The gap spacing between the actuation electrodes and the mechanical resonator is 150nm. The suspended waveguide is anchored to the substrate using tethers to prevent its collapse. S-bend structures are used to reduce the optical loss at these anchor points [19]. Additionally, the disk resonators have etch holes in them to allow the oxide etchant to undercut the disks leaving behind a 1μm diameter pedestal at the center. Grating couplers are used to couple light into and out of the waveguides with a loss of 10.8dB at each coupler.

To characterize the optical resonator, light from a tunable laser polarized parallel to the substrate is coupled into the waveguides using gratings. The optical transmission spectrum shows multiple dips corresponding to the whispering gallery modes of the disk resonator. An optical *Q* of 26,140 is obtained for a resonance at 1550.541 nm with an extinction of 10 dB as shown in figure 2. To excite mechanical excitations and observe the optical response, a setup as shown in figure 3a is used. The tunable laser is biased at the half maximum point of the optical resonance and a DC bias along with the RF output from port 1 of a network analyzer is applied to the actuation electrodes. The optical output from the gratings is picked up using an amplified high speed photodiode and sent to port 2 of the network analyzer.

The optical modulation at the mechanical resonant frequency is seen in the S21 plot of the network analyzer in figure 3c. The data show that modulation occurs only when the electrical input to the device is around the mechanical resonance frequency of the device. The response shows the presence of two modes of vibration at 235.1 MHz and 242.1 MHz due to splitting of the mechanical mode due to the coupling spring. The mode at 235.1 MHz has a mechanical quality factor of 1,300 in air. Increasing the DC voltage results in a larger actuation force and thereby an increase in amplitude of vibration. This leads to larger modulation of optical power for a given input RF power and is seen as a higher value of the transmission (S21). This confirms the hypothesis that the actuation is indeed caused by electrostatics. The predicted mechanical frequencies and the corresponding measured values are in good agreement with each other. In addition to the in phase and out of phase radial vibration modes, we also observe excitation of other vibration modes.

The amplitude of displacement of the disk is calculated by using the modulated power at the network analyzer input. Taking into account the losses, the amplitude of optical power modulation around the average optical power at the bias point for varying RF input and DC bias voltages is shown in figure 4. It is seen that the extinction ratio, defined as $10 log_{10}(P_{max}/P_{min})$ increases with increasing RF power and DC voltage and is a maximum of 6dB at 15V DC bias and 5 dBm RF power, corresponding to a radial displacement of 0.12nm. A larger modulation depth can be obtained for the same power and bias voltages by increasing the mechanical quality factor of the disks, which are currently limited by the anchor loss. As there is only one layer to fabricate the resonators and the electrical routing beams, the disk is electrically grounded by a beam that is connected to its edge. This beam is at the maximum displacement point of the radial mode of the disk increasing anchor loss leading to a lower $Q_{mech}$.

We have demonstrated a monolithic electrically actuated silicon optomechanical resonator operating at a frequency of 235MHz. There is no DC power dissipated in this device as the DC bias is applied across a capacitor. As the optical response to the electrical input is shaped by the mechanical transfer function, the device is similar to a RF filter in series with an optical modulator. This device can provide a low-power, on-chip modulator for narrow band applications such as a chip scale implementation of an Optoelectronic Oscillator [20, 21]. In its application as narrowband modulator, the device presented can remove inefficiencies associated with the repeated conversion from electrical → acoustic-filter → electrical → impedance-match → electrical → optical by directly going from electrical → acoustic-filter → optical. This device offers an additional input for controlling the state of an optomechanical system, and therefore has interesting applications in studying physics.

**Methods:**

The modulator is fabricated using a two mask process on a custom SOI wafer (undoped 250 nm device layer for low optical loss and 3μm thick buried oxide for isolation of the waveguides on device layer from the silicon substrate). The top silicon is thermally oxidized to obtain a thin oxide hard mask layer and a silicon device layer thickness of 220nm. Ma-N 2403 electron beam resist is spun on top of the oxide and patterned using electron beam lithography. The patterns are transferred into the oxide using a $CHF_3/O_2$ based reactive ion etcher and then into the silicon device layer using a chlorine based reactive ion etch to define the modulator, waveguides and bond-pads. A second layer of photoresist is spun and windows are patterned above the mechanical resonator and electrical routing beams. A boron ion implantation is carried out to reduce the resistivity of these structures. A third mask is then used to pattern release windows near the modulator followed by a timed release etch in buffered oxide etchant to undercut the devices (Figure 4). The samples are then dried using a critical point dryer to prevent stiction.

**Acknowledgements:**

The authors wish to thank the Semiconductor Optoelectronic Group and the Cornell Nanophotonics Group for helpful discussions. This work was supported by the DARPA Young Faculty Award and DARPA/MTO's ORCHID program and was carried out in part at the Cornell NanoScale Science and Technology Facility, which is supported by the National Science Foundation.

**Author Contributions:**

Suresh Sridaran and Sunil A. Bhave designed the device. Suresh Sridaran fabricated the device and performed experiments.

**Competing financial interests:** The authors declare no competing financial interests.

**Figure Legends**

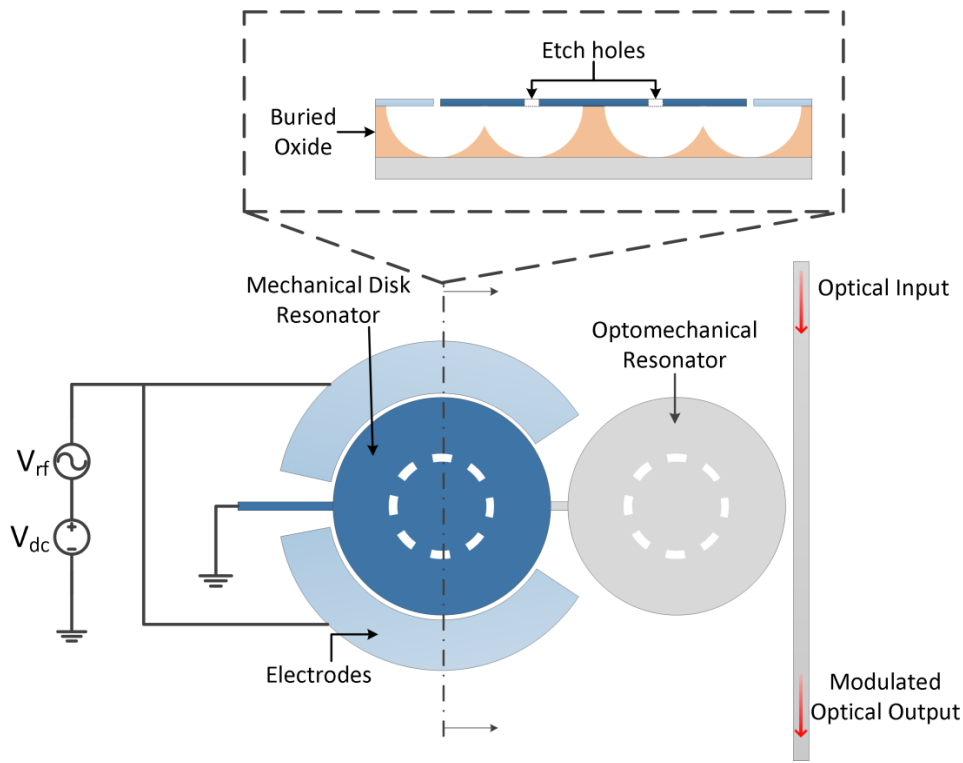

**Figure 1: Schematic of the electrically actuated disk mechanical resonator in blue (on left) coupled to an optomechanical resonator in grey (on right). A combination of DC and RF potential applied to electrodes around the mechanical resonator excite vibrations which get coupled to the optomechanical disk resonator .These radial vibrations affect the optical resonance thereby modulating the intensity of light exiting the waveguide.**

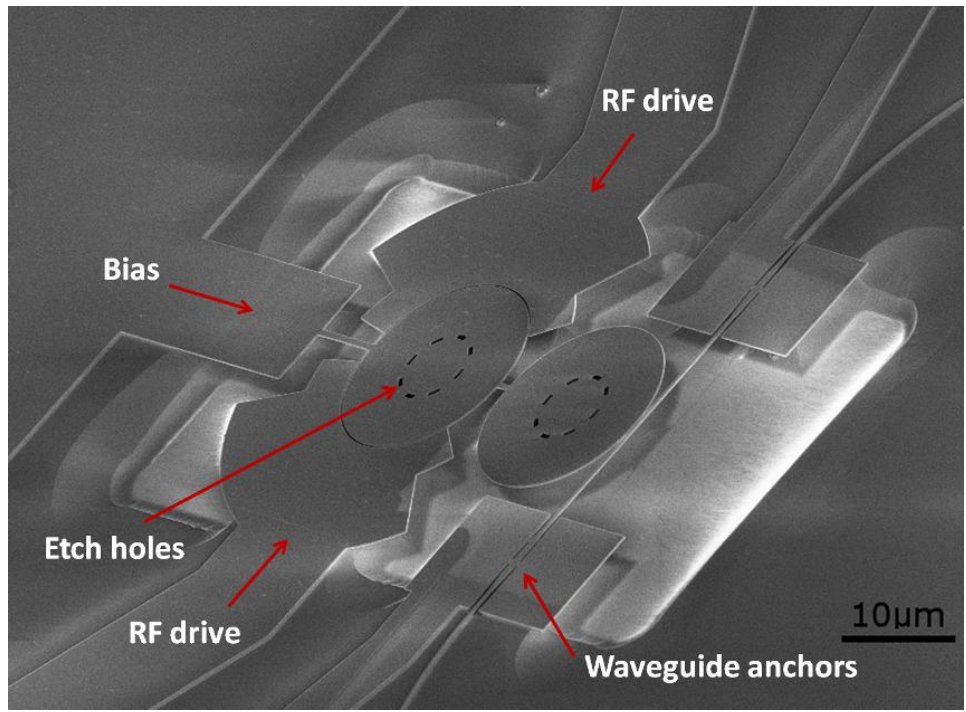

**Figure 2.** Scanning electron microscope image of fabricated device showing the coupled disk resonator geometry. The darker silicon shade show the silicon regions where boron has been ion implanted while the lighter silicon regions are undoped silicon.

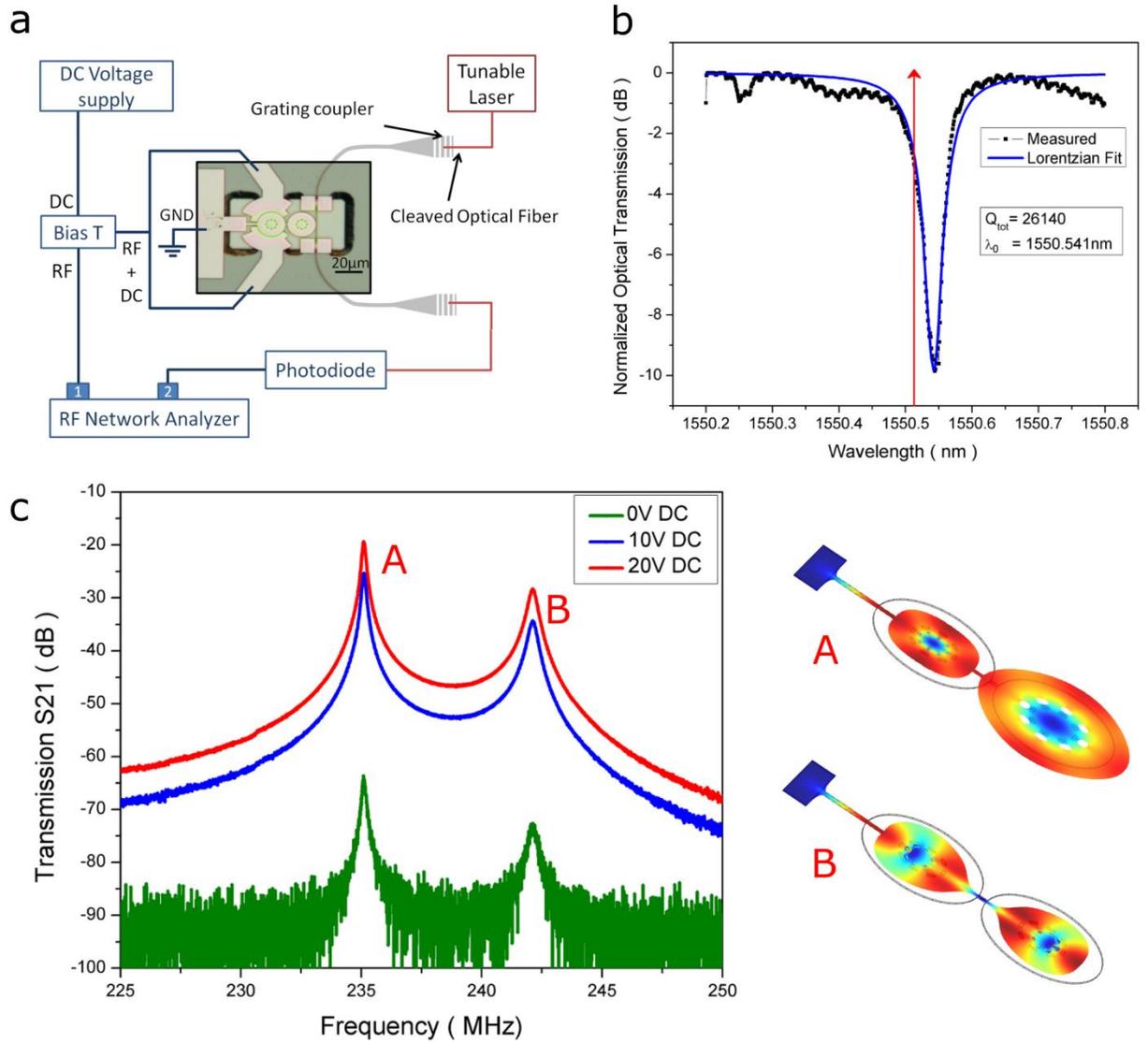

Figure 1a. Setup for measuring the optical response of electrically actuated optomechanical resonators. 3b. Optical transmission spectrum showing the disk resonance at 1550.541nm with a quality factor of 26,140. The red line shows the wavelength of the tunable laser for the modulator measurements   3c. Electrical transmission spectrum observed using the network analyzer showing modulation peaks at 235.1 MHz (A) and 242.1MHz (B) when input RF power is at -10dBm and optical input power is at 6dBm. Inset A and B show the displacement mode shape obtained using finite element method for modes predicted to be at 241.6MHz and 248.3MHz respectively

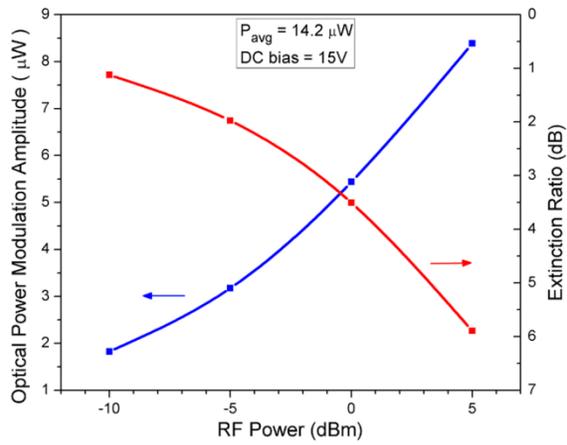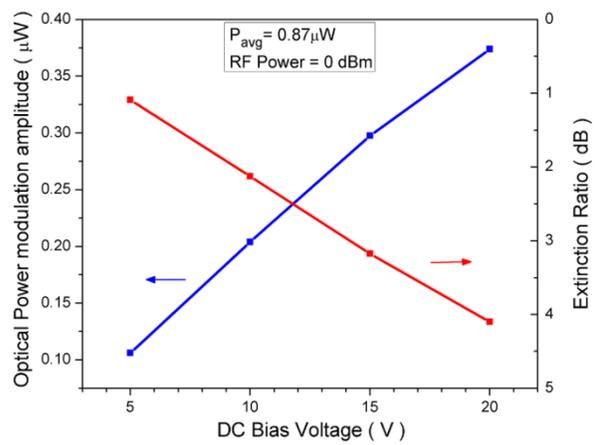

**Figure 4. Variation of the amplitude of optical power modulation and extinction ratio with RF power (left) and DC bias voltage (right) for an optical input power of 6dBm(left) and -6.0 dBm(right)**